\documentclass{jpsj3}
\usepackage{txfonts}
\usepackage{color}
\usepackage{graphicx}
\usepackage{amsmath}
\usepackage{amsfonts}
\title{Message-passing algorithm of quantum annealing\\ with nonstoquastic Hamiltonian}

\author{Masayuki Ohzeki\thanks{mohzeki@tohoku.ac.jp}}
\inst{Graduate School of Information Science, Tohoku University, Sendai, Japan \\
Institute of Innovative Research, Tokyo Institute of Technology, Kanagawa, Japan \\
Jij Inc., Tokyo, Japan}

\abst{
Quantum annealing (QA) is a generic method for solving optimization problems using fictitious quantum fluctuation.
The current device performing QA involves controlling the transverse field; it is classically simulatable by using the standard technique for mapping the quantum spin systems to the classical ones.
In this sense, the current system for QA is not powerful despite utilizing quantum fluctuation.
Hence, we developed a system with a time-dependent Hamiltonian consisting of a combination of the formulated Ising model and the ``driver" Hamiltonian with only quantum fluctuation.
In the previous study, for a fully connected spin model, quantum fluctuation can be addressed in a relatively simple way. 
We proved that the fully connected antiferromagnetic interaction can be transformed into a fluctuating transverse field and is thus classically simulatable at sufficiently low temperatures.
Using the fluctuating transverse field, we established several ways to simulate part of the nonstoquastic Hamiltonian on classical computers. 
We formulated a message-passing algorithm in the present study. 
This algorithm is capable of assessing the performance of QA with part of the nonstoquastic Hamiltonian having a large number of spins.
In other words, we developed a different approach for simulating the nonstoquastic Hamiltonian without using the quantum Monte Carlo technique.
Our results were validated by comparison to the results obtained by the replica method.
}

\begin{document}

\maketitle

\section{Introduction}
Quantum annealing (QA) is a generic algorithm aimed at solving optimization problems by exploiting the quantum tunneling effect.
The scheme was originally proposed as an algorithm for numerical computation \cite{Kadowaki1998} inspired by simulated annealing (SA) \cite{Kirkpatrick1983} and exchange Monte Carlo simulation \cite{Hukushima1996}.
However, its experimental realization was accomplished recently, attracting considerable attention.
QA is advantageous for solving an optimization problem formulated with discrete variables using a platform, the Ising model.
In QA, we developed a system with a time-dependent Hamiltonian consisting of a combination of the formulated Ising model and the ``driver" Hamiltonian with only quantum fluctuation.
The often-used driver Hamiltonian is the transverse field, which generates the superposition of the up and down spins.
The first stage of QA is initialized in the trivial ground state of the driver Hamiltonian.
The quantum effect will be gradually turned off and eventually ends so that only the formulated Ising model with the nontrivial ground state remains.
When the transverse field changes sufficiently slowly, the quantum adiabatic theorem ensures that we find the nontrivial ground state at the end of QA \cite{Suzuki2005,Morita2008,Ohzeki2011c}.
Numerous reports state that QA outperforms SA \cite{Santoro2002,Santoro2004,Baldassi2018}.
Possibly, the performance of QA stems from the quantum tunneling effect penetrating the valley of the potential energy.
The protocol of QA is realized in an actual quantum device using current technology, namely, quantum annealer \cite{Dwave2010a,Dwave2010b,Dwave2010c,Dwave2014}.
The output solution from the current system comprising quantum annealer is not always optimal due to device limitations and environmental effects \cite{Amin2015}.
Therefore, several protocols based on QA do not follow several conditions in adiabatic quantum computation or maintain a system in the ground state to reach the optimal solution in the final stage of QA; rather, they employ a nonadiabatic counterpart \cite{Ohzeki2010a,Ohzeki2011,Ohzeki2011proc,Somma2012}.
Nevertheless, quantum annealer has been tested for numerous applications such as portfolio optimization \cite{Rosenberg2016}, protein folding \cite{Perdomo2012}, the molecular similarity problem \cite{Hernandez2017}, computational biology \cite{Richard2018}, job-shop scheduling \cite{Venturelli2015}, traffic optimization \cite{Neukart2017}, election forecasting \cite{Henderson2018}, machine learning \cite{Crawford2016,Neukart2018,Khoshaman2018}, and automated guided vehicles in plants \cite{Ohzeki2019}.
In addition, studies on solving various problems by using quantum annealer have been performed \cite{Arai2018nn,Takahashi2018,Ohzeki2018NOLTA,Okada2019}. 
The potential of QA might be boosted by the nontrivial quantum fluctuation, referred to as the nonstoquastic Hamiltonian, for which efficient classical simulation is intractable \cite{Seki2012,Seki2015,Ohzeki2017,Arai2018dy}.

The efficiency of QA is characterized by the energy gap between the ground state and the first excited state in the intermediate Hamiltonian.
The adiabatic theorem formulates the relationship between the energy gap $\Delta$ and the necessary computation time $\tau_{\rm QA}$ for QA as $\tau_{\rm QA} \sim 1/\Delta^2$ \cite{Suzuki2005}.
The necessary computation time can be longer when the size of the problems, which is characterized by the number of spins $N$, increases because the energy gap decreases.
When the system is involved in phase transition, the energy gap collapses exponentially or polynomially.
The former is a first-order phase transition, and the latter is a second-order one.
Correspondingly, the order of the phase transition describes the performance of QA.
In particular, the existence of the first-order phase transition during hampers efficient manipulation of QA to find the nontrivial ground state at the end.
In this sense, avoiding the first-order phase transition is one of the central issues in QA.

One of the possible ways to avoid the first-order phase transition is to introduce the antiferromagnetic XX interaction into the conventional time-dependent Hamiltonian in QA \cite{Seki2012,Seki2015,Okada2019XX}.
In particular, the quantum fluctuation of the antiferromagnetic XX interaction is useful for eliminating the sudden change in the ferromagnetic ordering of the Ising spins.
Thus, the first-order phase transition can be modified into the second-order one.
One of the nontrivial properties of the antiferromagnetic XX interaction is the emergence of the negative sign in the standard way to map the quantum spin systems with the effective classical spin systems.
For instance, the Suzuki--Trotter decomposition \cite{Suzuki1976} yields a negative sign for the antiferromagnetic XX interaction in a straightforward way.
In this sense, the Ising spin models with antiferromagnetic XX interactions are nonstoquastic \cite{Bravyi2008}.
The nonstoquastic Hamiltonian is expected to enhance the power of quantum computation because it is capable of addressing intractable quantum fluctuation on a classical computer.
Thus, the antiferromagnetic XX interaction is also expected to be with certain potential for quantum speed up.
However, beyond modifying ferromagnetic ordering, the advantage of antiferromagnetic XX interactions is not yet clarified.

In the previous study \cite{Ohzeki2017}, they proved that the fully connected antiferromagnetic interaction can be transformed into a fluctuating transverse field and is thus classically simulatable at sufficiently low temperatures; this can be implemented in a system with a large number of spins using the quantum Monte Carlo simulation.
In the present study, we formulated the message-passing algorithm and its approximated version.
The resultant algorithm is capable of assessing the performance of QA with part of the nonstoquastic Hamiltonian having a large number of spins.

The remainder of the paper is organized as follows:
In the next section, we discuss the transformation of the antiferromagnetic XX interactions into a fluctuating transverse field and building of the quantum Monte Carlo simulation for part of the nonstoquastic Hamiltonian.
In the following section, we describe the message-passing algorithm for the effective model including the antiferromagnetic XX interactions.
We then discuss the asymptotic behavior for a large number of spins.
In the conclusion section, we summarize our findings and discuss future research directions.

\section{Transformation of Non-stoquastic Hamiltonian}
Let us consider the following quantum system for solving the optimization problem using QA.
The target Hamiltonian, which represents the optimization problem to be solved, can be written as
\begin{equation}
\hat{H}_0 = f(\hat{{\boldsymbol \sigma}}^z)
\end{equation}
where $\hat{{\boldsymbol \sigma}}^z = (\hat{\sigma}_1^z,\hat{\sigma}_2^z,\cdots,\hat{\sigma}_N^z)$ and $\hat{\sigma}_i^z$ is the $z$-component of the Pauli matrices.
The driver Hamiltonian is chosen with the following special form as
\begin{equation}
\hat{H}_1 = - N g\left( \frac{1}{N}\sum_{i=1}^N \hat{\sigma}_i^x \right).
\end{equation}
The function $g$ includes various types of quantum fluctuations.
For instance, the transverse field is the case with $g(x) = \Gamma x$ and the antiferromagnetic XX interaction, which is a typical example of the nonstoquastic Hamiltonian; that is, $g(x) = -\gamma x^2/2$.
Using the following decomposition, we can simulate part of the nonstoquastic Hamiltonian without facing the sign problem.
Let us consider the composite system comprising the target and driver Hamiltonians by computing the partition function.
\begin{equation}
Z = {\rm Tr}\left\{ \exp\left( - \beta \hat{H}_0 - \beta \hat{H}_1\right) \right\}
\end{equation}
Here, we employed the Suzuki--Trotter decomposition to separate the noncommutable operators in the exponential.
\begin{equation}
Z = \lim_{M \to \infty}{\rm Tr}\left\{ \prod_{k=1}^M \exp\left( - \frac{\beta}{M} f(\hat{{\boldsymbol \sigma}}_k^z)\right)\exp\left( \frac{N\beta}{M}g\left( \frac{1}{N}\sum_{i=1}^N \hat{\sigma}_{i,k}^x \right)\right) \right\}
\end{equation}
We may regard each operator in the Hamiltonian as the $c$ numbers after inserting the completeness relation $1=\sum_{{\boldsymbol \sigma}^z_k} |{\boldsymbol \sigma}^z_k\rangle \langle {\boldsymbol \sigma}^z_k|$ and $1=\sum_{{\boldsymbol \sigma}^x_k}|{\boldsymbol \sigma}^x_k \rangle \langle {\boldsymbol \sigma}^x_k|$.
In addition, we used the following identity and changed the delta function in terms of the Fourier integration as
\begin{equation}
1 =\int dm_{x,k} \delta( Nm_{x,k} - \sum_{i=1}^N \sigma_{i,k}^x) = \int d\tilde{m}_{x,k}\int dm_{x,k} \exp\left\{ - \frac{\beta \tilde{m}_{x,k}}{M}\left( Nm_{x,k} - \sum_{i=1}^N \sigma_{i,k}^x \right)\right\}
\end{equation}
After again considering the c numbers as the operators, the resultant expression of the partition function can be written as
\begin{equation}
Z \approx \lim_{M \to \infty}{\rm Tr}\left\{ \prod_{k=1}^M \int d\tilde{m}_{x,k}\int dm_{x,k} \exp\left(\frac{N\beta}{M}g\left(m_{x,k}\right) - \frac{N\beta}{M}\tilde{m}_{x,k}m_{x,k} \right) \exp\left( - \frac{\beta}{M} f(\hat{{\boldsymbol \sigma}}_k^z) \right) \exp\left( \frac{\beta\tilde{m}_{x,k}}{M} \sum_{i=1}^N\hat{\sigma}_{i,k}^x \right) \right\}.\label{first}
\end{equation}
Here, we utilized the inverse relation of the Suzuki--Trotter decomposition and obtained
\begin{equation}
Z \approx {\rm Tr}\left\{ \int d\tilde{m}_x \int dm_x\exp\left(N\beta g\left(m_{x}\right) - N\beta\tilde{m}_{x}m_{x} \right) \exp\left( - \beta f(\hat{{\boldsymbol \sigma}}^z) +  \beta\tilde{m}_{x} \sum_{i=1}^N\hat{\sigma}_{i}^x \right) \right\}.
\end{equation}
This expression led to the interpretation of the original composite system comprising the target and driver Hamiltonians as the following simple system with a fluctuating transverse field:
\begin{equation}
\hat{H}_{\rm eff} = f(\hat{{\boldsymbol \sigma}}^z) - \tilde{m}_{x} \sum_{i=1}^N\hat{\sigma}_{i}^x.
\end{equation}
The strength of the transverse field is governed by the probability distribution as follows:
\begin{equation}
P(\tilde{m}_x) \propto {\rm Tr}\left\{ \int dm_x\exp\left(N\beta g\left(m_{x}\right) - N\beta\tilde{m}_{x}m_{x} \right) \exp\left( - \beta f(\hat{{\boldsymbol \sigma}}^z) +  \beta\tilde{m}_{x} \sum_{i=1}^N\hat{\sigma}_{i}^x \right) \right\}.
\end{equation}
The resulting expression states that the nonstoquastic term $g(\sum_{i=1}^N \hat{\sigma}^x_i/N)$ can be reduced to the transverse-field term.
This fact implies that the present form of the nonstoquastic term does not represent strong quantum effects as the interference of the wave function.
However, the same type of nonstoquastic term changes the first-order phase transition, which is harmful to the second-order one and can be easily solved using QA \cite{Seki2012}. 

In QA, we were interested in extremely low-temperature regions as $\beta \to \infty$.
Therefore, we considered the saddle point.
The saddle point equations are given as
\begin{eqnarray}
m_{x} &=& \frac{1}{N} \sum_{i=1}^N {\rm Tr}\left( \hat{\rho}_{\rm eff} \hat{\sigma^x_{i}} \right) \label{sp0}\\
\tilde{m}_{x} &=& \beta g'\left( m_{x}\right).\label{sp2}
\end{eqnarray}
where $\hat{\rho}_{\rm eff} = \exp(-\beta \hat{H}_{\rm eff})/Z(\tilde{m}_x)$ and $Z(\tilde{m}_x) = {\rm Tr}\exp\left(-\beta \hat{H}_{\rm eff}\right)$.
Instead of the saddle-point equation, one may utilize the Langevin dynamics to generate the probability distribution.
However, it might take a long time to equilibrate the dynamics.
One would need to accelerate the equilibration as demonstrated in some studies \cite{Ichiki2013,Ohzeki2015,Ohzeki2015proc}.
After determining the transverse magnetization $m_x$ for a given value of $\tilde{m}_x$, we updated the value of $\tilde{m}_x$, which is the strength of the effective transverse magnetic field.
Iterative manipulation of the above procedure yields the equilibrium state of the system.
We can then compute the physical quantity in the equilibrium state.
In the current system implementing QA in the physical device using D-Wave 2000Q, the transverse field can be tuned by following the predetermined schedule.
The schedule allows for pausing the control of the transverse field in the intermediate period of the QA protocol.
We can estimate the physical quantities in the equilibrium state with the transverse field, as explained in a previous study \cite{Dwave2018}.
Nevertheless, unfortunately, the current system is not capable of estimating the transverse magnetization directly.
Therefore, we did not directly perform the above procedure using D-Wave 2000Q.
Instead of the quantum device, we may employ a classical simulation to estimate the transverse magnetization, similar to a previous study \cite{Ohzeki2017}.
They utilized the Suzuki--Trotter decomposition and employed the following relationship:
\begin{equation}
\left\langle {\boldsymbol \sigma}_{k+1}^z \right| \exp\left( \frac{\beta\tilde{m}_{x,k}}{M} \hat{\sigma}_{i,k}^x \right) \left|{\boldsymbol \sigma}_k^z \right\rangle \propto \left\langle {\boldsymbol \sigma}_{k+1}^z \right| \exp\left( \beta J^*_k \sum_{i=1}^N\sigma_{i,k}^z\sigma_{i,k+1}^z \right) \left|{\boldsymbol \sigma}_k^z \right\rangle
\end{equation}
where $\exp\left(-2\beta J^*\right) = \tanh \beta \tilde{m}_{x,k}/M$.
In other words, the original quantum system can be mapped onto a classical effective Hamiltonian with spin variables $\sigma_{i,k}^z = \pm 1$ and continuous variables $\tilde{m}_{x,k}$ and $m_{x,k}$.
Instead of using Eq. (\ref{sp0}), we estimated the transverse magnetization using the following relationship, which was straightforwardly obtained by taking the derivative with respect to $\tilde{m}_{x,k}$,
\begin{eqnarray}
m_{x,k} &=& \frac{1}{N} \sum_{i=1}^N \left\langle \left(\tanh  \frac{\beta \tilde{m}_{x,k}}{M} \right)^{\sigma^z_{i,k}\sigma^z_{i,k+1}}\right \rangle \label{sp1}
\end{eqnarray}
We should emphasize that the target Hamiltonian is not restricted to the case with all-to-all connections.
The target Hamiltonian can only be used with short-range interactions such as those in the finite-dimensional Edwards--Anderson model.

In the present study, we formulated the belief-propagation method to reduce the computational cost \cite{Yedida2003,Mezard2009}.
In addition, we derived an approximate message-passing algorithm, which led to results consistent with the analytical solution at the level of replica symmetry.
\section{Message-passing algorithm}
We hereafter assume that the typical spin-glass model takes the following form:
\begin{equation}
f({\boldsymbol \sigma}) = - \sum_{\mu} J_{\mu} \prod_{l \in \partial \mu} \sigma_l -\sum_{i=1} h_i \sigma_i.
\end{equation}
where $\mu$ denotes the factor nodes to which the interactions between spins are assigned, and $k$ denotes variable nodes to which the spin variables are allocated.
This assumption is reasonable for the standard definition of spin-glass models.

In the classical case without the transverse field, we iteratively computed the following update equation:
\begin{eqnarray}
M_{\mu \to i}(\sigma_i) \propto \sum_{{\boldsymbol \sigma}/\sigma_i}\exp\left(\beta J_{\mu}\prod_{l \in \partial \mu}\sigma_{l}\right) \prod_{l \in \partial \mu/i}M_{l \to \mu}(\sigma_l) \label{msg1}\\
M_{i \to \mu}(\sigma_i)\propto \exp\left( \beta h_i \sigma_i\right) \prod_{\nu \in \partial i/\mu}M_{\nu \to i}(\sigma_i), \label{msg2}
\end{eqnarray}
where $M_{\mu \to i}(\sigma_i)$ and $M_{i \to \mu}(\sigma_i)$ are the messages passing through the factor and variable nodes, respectively.
The notation ${\boldsymbol \sigma}/\sigma_i$ denotes that the summation over the  variables except for $\sigma_i$.
These messages form the approximation of the distribution function through
\begin{eqnarray}
Q_{\mu}({\boldsymbol \sigma}_{\partial \mu}) \propto \sum_{\boldsymbol \sigma}\exp\left(\beta J_{\mu}\prod_{i \in \partial \mu} \sigma_{\mu}\right)\prod_{i \in \partial \mu} M_{i \to \mu}(\sigma_i)  \\
Q_{i}(\sigma_i) \propto  \exp\left( \beta h_i \sigma_i\right)  \prod_{\mu \in \partial i}M_{\mu \to i}(\sigma_i).
\end{eqnarray}
These ingredients approximate the distribution function according to the following relationship:
\begin{equation}
P({\boldsymbol \sigma}) \approx \prod_{\mu}\frac{Q_{\mu}({\boldsymbol \sigma}_{\partial \mu})}{\prod_{i \in \partial \mu}Q_i(\sigma_i)} \prod_{i=1}^N Q_i(\sigma_i).
\end{equation}
This approximation can be obtained by minimization of the Kullback--Liebler divergence under constraints such as $\sum_{{\boldsymbol \sigma}_{\partial \mu/i}}Q_{\mu}({\boldsymbol \sigma}_{\partial \mu}) = Q_i(\sigma_i)$ and normalization.
The optimal solution satisfies the previous equations for messages (\ref{msg1}) and (\ref{msg2}).

Let the message be parametrized as 
\begin{eqnarray}
M_{\mu \to i}(\sigma_i) &\propto& \exp\left( \beta \tilde{m}_{\mu \to i}\sigma_i \right) \\
M_{i \to \mu}(\sigma_i) &\propto& \exp\left( \beta \tilde{m}_{i \to \mu}\sigma_i \right).
\end{eqnarray}
These parametrizations are validated by the fact that the message is a function of the local Ising spin variable.
We obtained the first update equation as
\begin{equation}
\tilde{m}_{i \to \mu} = h_{i} + \sum_{\nu \in \partial i/\mu} \tilde{m}_{\nu \to i}\label{ueq1}
\end{equation}
The second update equation was determined by taking the summation over the Ising variable.
The message was assessed as follows:
\begin{equation}
M_{\mu \to i}(\sigma_i) \propto \cosh\left(\beta J_{\mu}\right) + \sigma_i \prod_{l \in \partial \mu}\left( \sum_{\sigma_l }M_{l \to \mu}(\sigma_l)\sigma_{l} \right)\sinh\left(\beta J_{\mu}\right).
\end{equation}
Thus, we obtained
\begin{eqnarray}
\tanh \beta \tilde{m}_{\mu \to i} = \tanh(\beta J_{\mu}) \prod_{l \in \partial \mu/i} m_{l \to \mu}.\label{ueq2}
\end{eqnarray}
We define the following quantity for convenience:
\begin{equation}
m_{l \to \mu} \equiv \frac{1}{Z_{l \to \mu}}\sum_{\sigma_l} \sigma_l  \exp\left( \beta \tilde{m}_{l \to \mu}\sigma_l\right) = \tanh\left( \beta \tilde{m}_{l \to \mu} \right).
\end{equation}
where $Z_{l\to \mu} = \sum_{\sigma_l}  \exp\left( \beta \tilde{m}_{l \to \mu}\sigma_l \right)$.
In other words, $m_{l \to \mu}$ is the local expectation of the Ising spin variable.
Using these messages, we obtained the following relationships:
\begin{eqnarray}
m_{i \to \mu} &=& \tanh\left( \beta \tilde{m}_{i \to \mu} \right) \\
\tilde{m}_{i \to \mu} &=& h_i + \frac{1}{\beta} \sum_{\nu \in \partial i/\mu}\tanh^{-1}\left(\tanh \beta J_{\nu} \prod_{l \in \partial \nu/i} m_{l \to \nu} \right).\label{ueq2}
\end{eqnarray}
These equations are well-known results for the belief propagation based on the Ising spin-glass model.
For the local magnetization, we obtained
\begin{eqnarray}
m_{i} &=& \tanh\left( \beta \tilde{m}_{i} \right)\label{op0} \\
\tilde{m}_{i} &=& h_i + \frac{1}{\beta} \sum_{\nu \in \partial i}\tanh^{-1}\left(\tanh \beta J_{\nu} \prod_{l \in \partial \nu/i} m_{l \to \nu} \right).
\end{eqnarray}
Considering the above calculations, we found that the essential calculation was the expectation of the Ising spin variable following the local distribution characterized by $\tilde{m}_{l \to \mu}$.
We considered generalization of the local distribution to include the quantum fluctuations.
One of the simplest approaches is changing the local distribution $M_{i \to \mu}(\sigma_i)$ into the density matrix following the prescription of quantum mechanics as shown below:
\begin{eqnarray}
M_{i \to \mu}(\sigma_i) \propto \exp\left(\beta \tilde{m}_{i \to \mu}\sigma_i \right) \to
\hat{M}_{i \to \mu}  \propto \exp\left(\beta \tilde{m}_{i \to \mu}\hat{\sigma}_i^z + \beta \tilde{m}_i^x \hat{\sigma}_i^x \right).
\end{eqnarray}
We performed the Suzuki--Trotter decomposition of this density matrix and further generalized the message as long as we could construct the closed update equation as follows:
\begin{eqnarray}
M_{i \to \mu}({\boldsymbol \sigma}) \propto \exp\left(
\beta\tilde{m}_{i \to \mu}\sum_{k=1}^M \sigma_{i,k} + \frac{\beta}{2} \tilde{r}_{i \to \mu}\left(\sum_{k=1}^M\sigma_{i,k}\right)^2 +  \beta J_i^* \sigma_{i,k}\sigma_{i,k+1} \right).
\end{eqnarray}
This is rewritten using the Hubbard--Stratonovich transformation as follows:
\begin{eqnarray}
M_{i \to \mu}({\boldsymbol \sigma}) = \frac{1}{Y_{i \to \mu}\left(\tilde{m}_{i \to \mu},\tilde{r}_{i \to \mu},\tilde{m}_{i}^x\right)} \int Dy \prod_{k=1}^M \exp\left(
\beta \left(\tilde{m}_{i \to \mu} + \sqrt{\tilde{r}_{i \to \mu}/\beta}y\right)\sigma_{i,k} + \beta J_i^* \sigma_{i,k}\sigma_{i,k+1} \right),
\end{eqnarray}
where
\begin{equation}
Y_{i \to \mu} \left(\tilde{m}_{i \to \mu},\tilde{r}_{i \to \mu},\tilde{m}_{i}^x\right) = \sum_{\boldsymbol \sigma} \int Dy \prod_{k=1}^M \exp\left(
\beta \left(\tilde{m}_{i \to \mu} + \sqrt{\tilde{r}_{i \to \mu}/\beta}y\right)\sigma_{i,k} + \beta J_i^* \sigma_{i,k}\sigma_{i,k+1} \right),
\end{equation}
and $\int Dy = \int^{\infty}_{-\infty} dy \exp\left( -y^2/2\right)/\sqrt{2\pi}$.
Using the above form of the message, we did not always obtain the closed update equations.
When we omit higher-order terms other than $\beta J_{\mu}^2$, we obtain the following closed update equations.
For the longitudinal magnetization, we have Eq. (\ref{ueq2}) and 
\begin{equation}
m_{i \to \mu} = \frac{1}{M\beta}\frac{\partial}{\partial \tilde{m}_{i \to \mu}} \log Y_{i \to \mu} \left(\tilde{m}_{i \to \mu},\tilde{r}_{i \to \mu},\tilde{m}_{i}^x\right).\label{ueq3}
\end{equation}
In addition, we determined another set of update equations as follows:
\begin{eqnarray}
\tilde{r}_{i \to \mu} &=& \frac{1}{\beta} \sum_{\nu \in \partial i/\mu}\tanh^{-1}\left(\tanh^2 \beta J_{\nu} \left(\prod_{l \in \partial \nu/i} r_{l \to \mu}-\prod_{l \in \partial \nu/i} m^2_{l \to \mu}\right) \right)
 \label{ueq4}\\
r_{i \to \mu} &=& \frac{1}{M^2\beta}\frac{\partial}{\partial \tilde{r}_{i \to \mu}} \log Y_{i \to \mu} \left(\tilde{m}_{i \to \mu},\tilde{r}_{i \to \mu},\tilde{m}_{i}^x\right) \label{ueq5}
\end{eqnarray}
We then estimated both the longitudinal and transverse local magnetization to be
\begin{eqnarray}
m_{i} &=& \frac{1}{M\beta}\frac{\partial}{\partial \tilde{m}_{i \to \mu}} \log Y_{i \to \mu} \left(\tilde{m}_{i},\tilde{r}_{i},\tilde{m}_{i}^x\right)
\label{op1} \\
m^x_{i} &=& \frac{1}{M\beta}\frac{\partial}{\partial \tilde{m}_{i}^x} \log Y_{i \to \mu} \left(\tilde{m}_{i},\tilde{r}_{i},\tilde{m}_{i}^x\right)
\label{op2} \\
r_{i} &=&\frac{1}{M^2\beta}\frac{\partial}{\partial \tilde{r}_{i \to \mu}} \log Y_{i \to \mu} \left(\tilde{m}_{i \to \mu},\tilde{r}_{i \to \mu},\tilde{m}_{i}^x\right)
\label{op3}.
\end{eqnarray}
where 
\begin{eqnarray}
\tilde{r}_{i} &=&  \frac{1}{\beta} \sum_{\nu \in \partial i}\tanh^{-1}\left(\tanh^2 \beta J_{\nu} \left(\prod_{l \in \partial \nu/i} r_{l \to \mu} -\prod_{l \in \partial \nu/i} m^2_{l \to \mu} \right)\right).
\end{eqnarray}
We iterated the update equations for the messages in Eqs. (\ref{ueq2}), (\ref{ueq3}), (\ref{ueq4}), and (\ref{ueq5}) and let them converge to fixed-point solutions.
For $\tilde{m}_i^x$, we utilized the saddle-point equations (\ref{sp2}).
Then, we assessed $Y_{i \to \mu}\left(\tilde{m}_{i \to \mu},\tilde{r}_{i \to \mu},\tilde{m}_{i}^x\right)$.
It is possible to straightforwardly manipulate the calculation as
\begin{equation}
Y_{i \to \mu}\left(\tilde{m}_{i \to \mu},\tilde{r}_{i \to \mu},\tilde{m}_{i}^x\right)
= \int Dy 2\cosh\left( \beta\sqrt{(\tilde{m}_{i \to \mu} + \sqrt{\tilde{r}_{i \to \mu}/\beta} y )^2+ (\tilde{m}_{i}^x)^2}\right).
\end{equation}
The derived update equations allow us to elucidate the nontrivial results at the level of static approximation as shown below.
Notice that it is just approximation in terms of the belief propagation at the level of the Bethe approximation in terms of statistical mechanics.
We may use the above update equation for an arbitrary shape of the graph as in the loopy belief propagation, which is sometimes applied in signal processing and image restoration.
Similar to the case of the loopy belief propagation, we may encounter some nonconvergent behaviors depending on the problem.

We might generalize the message in a different form.
For instance, one of the candidates is
\begin{eqnarray}
M_{i \to \mu}({\boldsymbol \sigma}) \propto \prod_{k=1}^M \exp\left(
\beta \left(\tilde{m}_{i \to \mu,k}\sigma_{i,k} + J_i^* \sigma_{i,k}\sigma_{i,k+1}\right) \right).
\end{eqnarray}
This is the same computation for assessing the normalization constant as that for the partition function of the one-dimensional Ising model under the ``random field" defined by $\tilde{m}_{i \to \mu,k}$.
Using the transfer matrix method, we can compute the normalization constant \cite{Krzakala2008}.
This approach corresponds to the path-integral representation in the quantum cavity method according to literature.
In this sense, the form of the message is essentially important to characterize the level of the approximation of the approach.
To obtain the nontrivial results beyond the static approximation, we have to tackle a generic form of the normalization constant \cite{Giulio2001}.
In the present study, we analyzed the effect of the nonstoquastic Hamiltonian at the level of the static approximation as shown below.
Several terms, $\tilde{m}_{i \to \mu}$ and $\tilde{r}_{i\to \mu}$, consist of the summation of the random variables.
When we consider $|\partial i| \sim N \to \infty$, namely the fully connected models, according to the central limit theorem, $\tilde{m}_{i \to \mu}$ and $\tilde{r}_{i \to \mu}$ can be characterized by their mean and covariance.
As discussed below, the above message-passing algorithm for the fully connected models can be validated at the level of the replica symmetric and static approximation.

\section{Validation of our results}
\subsection{p-body spin-glass model}
We assumed that the interaction term covering all the combinations of the Ising variables and their strength scales as $J_{\mu} \sim O(1/N^{(p-1)/2})$, where $p$ is the number of the Ising spin variables involved in the factor $\mu$.
Then, $\tanh(\beta J_{\mu}) \approx \beta J_{\mu}$ and $\tanh^{-1}( \beta J_{\mu} \prod_{l \in \partial \nu/i}m_{l \to \nu}) \approx \beta J_{\mu} \prod_{l \in \partial \nu/i}m_{l \to \nu}$.
In a large-sized limit such as $N \to \infty$, the self-averaging property assures that $J_{\mu}$ and $h_i$ are random variables following their distribution function.
The update equation of $\tilde{m}_{i \to \mu}$ for the case without the transverse field is reduced to
\begin{equation}
\tilde{m}_{i \to \mu} = h_i + \frac{1}{\beta}\sum_{\nu \in \partial i/\mu}\beta J_{\nu} \prod_{l \in \partial \nu/i}m_{l \to \nu}.
\end{equation}
Similarly, $\tilde{m}_i$ is given by
\begin{equation}
\tilde{m}_{i} = h_i + \frac{1}{\beta}\sum_{\nu \in \partial i}\beta J_{\nu} \prod_{l \in \partial \nu/i}m_{l \to \nu}.
\end{equation}
The local magnetization and message have the following relationship:
\begin{equation}
m_{i \to \mu} = m_i - (1-m_i^2)\beta J_{\mu} \prod_{l \in \partial \mu/i}m_{l \to \mu}.
\end{equation}
The local magnetization $m_i = \tanh(\beta m_i)$ is thus given by, though Eq. (\ref{op0}), 
\begin{equation}
\tilde{m}_{i} = h_i + \sum_{\nu \in \partial i} J_{\nu} \prod_{l \in \partial \nu/i} m_{l} - \prod_{l \in \partial \nu/i}(1-m_l^2)\beta J_{\nu} \prod_{n \in \partial \nu/l}m_{n}.
\end{equation}
This is the Thouless--Anderson—Palmer (TAP) equation \cite{TAP1977,Opper2001}.
For the quenched interaction and field, the local magnetization can fluctuate over the site $i$.
We then estimated the local magnetization through the TAP equation.
On the other hand, the dynamics of the expectation value of the local magnetization over the site dependence can be assessed using the central limit theorem. 
We computed the expectation and variance of $\tilde{m}_i$ as
\begin{eqnarray}
{\mathbb E}\left[ \tilde{m}_i \right] &=& h \\
{\mathbb E}\left[ \tilde{m}_i^2 \right] - \left({\mathbb E}\left[ \tilde{m}_i \right]\right)^2 &=& J^2 q^{p-1}
\end{eqnarray}
where ${\mathbb E}\left[ \cdots \right]$ is the expectation on the probability distribution of $J_{\mu}$ with a mean equal to $J_0/N^{p-1}$ and variance equal to $\sqrt{J^2/N^{p-1}}$; we define the magnetization as $m = \sum_{i=1}^Nm_i/N$ and the spin-glass order parameter as $q=\sum_{i=1}^N m_i^2/N$.
Here, for simplicity, we set the magnetic field as constant.
Then, $\tilde{m}_i$ becomes 
\begin{equation}
\tilde{m}_i = h + J_0 p m^{p-1} + \sqrt{J^2p q^{p-1}}z_i
\end{equation}
where $z_i$ is the random variable with zero mean and unit variance.
Therefore, the magnetization can be evaluated as follows:
\begin{equation}
m = \int Dz \tanh\left\{ \beta\left(h + J_0 p m^{p-1} + \sqrt{J^2 p q^{p-1}}z\right)\right\}
\end{equation}
Similar to the magnetization, we obtained the spin-glass order parameter as follows:
\begin{equation}
q = \int Dz \tanh^2\left\{ \beta\left(h + J_0 p m^{p-1} + \sqrt{J^2p q^{p-1}}z\right)\right\}.
\end{equation}
These results are consistent with the result of the replica method for the $p$-body spin-glass model under the replica symmetric assumption.

Next, we discuss the quantum system.
We again computed the expectations and variances of $\tilde{m}_i$ and $\tilde{r}_i$ as
\begin{eqnarray}
{\mathbb E}\left[ \tilde{m}_{i} \right] &=& h + J_0 p m^{p-1} \\
{\mathbb E}\left[ \tilde{r}_{i} \right] &=& \beta J^2 p \left(R^{p-1}-q^{p-1}\right) \\
{\mathbb E}\left[ \tilde{m}_{i}^2 \right] - \left({\mathbb E}\left[ \tilde{m}_{i} \right]\right)^2 &=& J^2 p q^{p-1}
\end{eqnarray}
where we define the magnetization as $m = \sum_{i=1}^Nm_{i}/N$, the spin-glass order parameter as $q=\sum_{i=1}^N m_{i}^2/N$, and an overlap between different imaginary times as $R=\sum_{i=1}^N r_{i}/N$.
Then, we obtain
\begin{equation}
\tilde{m}_{i} = h + J_0 p m^{p-1} + \sqrt{pJ^2q^{p-1}}z_i.
\end{equation}
We set $a(y,z) \equiv h + J_0 p m^{p-1} + \sqrt{pJ^2q^{p-1}}z +  \sqrt{pJ^2(R^{p-1}-q^{p-1})}y$.
Using these results, we can reproduce the update equations, which are consistent with the saddle-point equations for the $p$-body spin-glass model under a uniform transverse field $\tilde{m}_i^x = \tilde{m}_x$ \cite{Obuchi2007} as follows: 
\begin{eqnarray}
m &=& \int Dz \frac{1}{Y_{i \to \mu}} \int Dy \frac{a(y,z)}{\sqrt{a(y,z)^2+(\tilde{m}^x)^2}}\sinh \beta\sqrt{a(y,z)^2+(\tilde{m}^x)^2}\label{SE1}
 \\ \nonumber
R &=& \int Dz \frac{1}{Y_{i \to \mu}} \int Dy \left(\frac{a(y,z)}{\sqrt{a(y,z)^2+(\tilde{m}^x)^2}}\right)^2\sinh \beta\sqrt{a(y,z)^2+(\tilde{m}^x)^2}\\
& &  + \int Dz \frac{1}{Y_{i \to \mu}}\int Dy \frac{\tilde{m}^x}{(a(y,z)^2+(\tilde{m}^x)^2)^{3/2}} \cosh \beta\sqrt{a(y,z)^2+(\tilde{m}^x)^2}\label{SE2}\\
q &=& \int Dz \left(\frac{1}{Y_{i \to \mu}}\int Dy \frac{a(y,z)}{\sqrt{a(y,z)^2+(\tilde{m}^x)^2}}\sinh \beta\sqrt{a(y,z)^2+(\tilde{m}^x)^2} \right)^2 \label{SE3}\\
m^x &=&\int Dz \frac{1}{Y_{i \to \mu}} \int Dy \frac{m^x}{\sqrt{a(y,z)^2+(\tilde{m}^x)^2}}\sinh \beta\sqrt{a(y,z)^2+(\tilde{m}^x)^2},\label{SE4}
\end{eqnarray}
where we omit the arguments of the function $Y_{i \to \mu}$ to lighten the writing.
Instead of the quantum Monte Carlo simulation, we employed the above update equations for estimating the order parameters conditioned on $\tilde{m}^x$.
The effective magnetic field changes according to the saddle-point equation (\ref{sp2}).
Iterative manipulation of these procedures yields the results for the equilibrium state of the fully connected $p$-body Ising model with antiferromagnetic XX interactions.
The results are generalizations of those obtained in previous studies \cite{Seki2012,Seki2015}.
In other words, we developed a different approach for the nonstoquastic Hamiltonian without using the quantum Monte Carlo simulations and the replica method.

\subsection{Application to restricted Boltzmann machine}
We introduce an application of our message passing algorithm for a part of the non-stoquastic Hamiltonian.
We take the restricted Boltzmann machine, whose Hamiltonian is given as
\begin{equation}
f({\bf v},{\bf h}) = - \sum_{i=1}^{N_v} b^v_i v_i - \sum_{j=1}^{N_h} b^h_j h_j - \sum_{i,j}W_{ij}v_ih_j.
\end{equation}
where $v_i \in {-1,1}$ and $h_j \in {-1,1}$ are the visible and hidden variables, respectively.
We regard these variables as the $z$-components of the Pauli matrices.
We then apply the transverse field and antiferromagnetic XX interactions for each variable.
We construct the message passing algorithm by defining two sets of the messages for visible and hidden variables.
\begin{eqnarray}
M_{\mu \to i}(v_i) &\propto& \sum_{h_j}\exp\left(\beta W_{\mu}h_j \right) M_{j \to \mu}(h_j) \\
M_{\mu \to j}(h_j) &\propto& \sum_{v_i}\exp\left(\beta W_{\mu}v_i \right) M_{i \to \mu}(v_i)
\end{eqnarray}
where $\mu = (ij)$, $M_{i \to \mu}$, and $M_{j \to \mu}$ similarly to Eq. (\ref{msg2}).
The direct manipulation of the above calculation yields the parameters of the messages as
\begin{eqnarray}
\tilde{m}_{i \to \mu} &=& b^v_i + \frac{1}{\beta} \sum_{j}\tanh^{-1}\left(\tanh \beta W_{\mu} m_{j \to \mu} \right) \\
\tilde{r}_{i \to \mu} &=& \frac{1}{\beta} \sum_{j}\tanh^{-1}\left(\tanh^2 \beta W_{\mu} \left( r_{j \to \mu}-m^2_{j \to \mu}\right) \right)
\end{eqnarray}
for the visible variables and similarly for the hidden variables by replacement of $i \leftrightarrow j$ and $b^v_i \leftrightarrow b^h_j$.
We find that $m_{i \to \mu}$, $m_{j \to \mu}$, $r_{i \to \mu}$, and $r_{j \to \mu}$ follow the same forms as Eqs. (\ref{ueq3}) and (\ref{ueq5}).
Both the longitudinal and transverse local magnetization can be obtained by Eqs. (\ref{op1}), (\ref{op2}) and (\ref{op3}).
The numbers of the visible and hidden variables are $N_v$ and $N_h$.
We assume $N_v$ and $N_v$ go infinity while $\alpha_v = N_v/(N_v+N_h)$ and $\alpha_h = N_h/(N_v+N_h)$ are finite.
Then we obtain, by the similar way to the previous case,
\begin{eqnarray}
\tilde{m}_i &=& b_i^v + \sum_{j} W_{\mu} m_j - (1-m_j)^2 \beta W_{\mu} m_i \\
\tilde{r}_i &=& \beta \sum_{j} W^2_{\mu}\left(r_j - m_j^2\right).
\end{eqnarray}
for the visible variables and similarly for the hidden variables by replacement of $i \leftrightarrow j$ and $b^v_i \leftrightarrow b^h_j$.
The restricted Boltzmann machine generates many samples in order to calculate the expectation value and variance in learning.
However our message-passing algorithm make sampling easier and faster because we omit relaxation time to generate the Gibbs-Boltzmann sampling, which needs relatively longer time.
In order to validate our result, we assume that $\mathbb{E}[W_{\mu}] = w/\sqrt{N_vN_h}$ and $\mathbb{E}[W_{\mu}^2] = W^2/\sqrt{N_vN_h}$ and no biases exist for simplicity.
The expectations and variances of $\tilde{m}_i$ and $\tilde{m}_j$ are
\begin{eqnarray}
{\mathbb E}\left[ \tilde{m}_{i} \right] &=& \sqrt{\frac{\alpha_v}{\alpha_h}} w m_h \\
{\mathbb E}\left[ \tilde{r}_{i} \right] &=& \beta \sqrt{\frac{\alpha_v}{\alpha_h}} W^2 \left(R_h-q_h\right) \\
{\mathbb E}\left[ \tilde{m}_{i}^2 \right] - \left({\mathbb E}\left[ \tilde{m}_{i} \right]\right)^2 &=& \sqrt{\frac{\alpha_v}{\alpha_h}}W^2 q_h
\end{eqnarray}
for the visible variables and similarly for the hidden variables by replacement of $i \leftrightarrow j$ and $(b_v, B_v, m_v, r_v, q_v) \leftrightarrow (b_h, B_h, m_h, r_h, q_h)$.
Then $\tilde{m}_{i}$ is expressed by, according to the central limit theorem,
\begin{equation}
\tilde{m}_i = \sqrt{\frac{\alpha_v}{\alpha_h}} w m_h + \left(\frac{\alpha_v}{\alpha_h}\right)^{1/4}W \sqrt{q_h}z
\end{equation}
Replacement of $a(y,z) = \sqrt{\alpha_v/\alpha_h} w m_h + \left(\alpha_v/\alpha_h\right)^{1/4}W \sqrt{q_h} z + \left(\alpha_v/\alpha_h\right)^{1/4} W \sqrt{R_h-q_h}y$ in Eqs. (\ref{SE1}), (\ref{SE2}), (\ref{SE3}) and (\ref{SE4}) leads to a natural generalization of the results of the previous study \cite{Hartnett2018}.
We emphasize that our results include the quantum fluctuation expressed by the transverse field and beyond.
In the restricted Boltzmann machine, we need many samples to estimate the expectation and variance in learning.
When we add the quantum fluctuation to the system, it is a harmful computation to generate many samples through the quantum Monte-Carlo simulation.
Instead, we here propose the message-passing algorithm in a deterministic way and it mitigate the difficulty in sampling.
The results are validated by comparison with the natural generalization into the case with the quantum fluctuation.
As far as our knowledge, this is the first investigation of the quantum effect in the restricted Boltzmann machine in theory while several algorithms for learning are proposed \cite{misha2011,Amin2018}.
The detailed analysis on the restricted Boltzmann machine in the transverse field and beyond will be reported elsewhere.

We again emphasize that the proposed method is not restricted to the case of a fully connected Ising model.
At the level of belief propagation, we estimated both the longitudinal and transverse local magnetizations.
For obtaining more accurate results, we may employ the cluster variational method \cite{Kikuchi1951} and its variants \cite{Chertkov2006,Rizzo2010,Xiao2011,Ohzeki2013}.

\section{Summary}
We propose a message-passing algorithm for part of the nonstoquastic Hamiltonian such as the Ising model with antiferromagnetic XX interactions.
Previous studies focused on the adaptive quantum Monte Carlo simulation, which generates the spin configuration in the equilibrium state of the limited range of the nonstoquastic Hamiltonian.
However, it is often the case with long-time equilibration.
Thus, belief propagation, despite being just an approximation, is valuable for investigating the equilibrium state of the nonstoquastic Hamiltonian.
One of the nontrivial features in the nonstoquastic Hamiltonian is mitigating the computational complexity in QA by avoiding the first-order phase transition.
This phenomenon is just a special case of several spin models showing a remarkable improvement in the performance of QA.
Although such a nontrivial feature is not universal when using the nonstoquastic Hamiltonian, it is a significant step toward understanding the potential of QA.
However, avoiding the first-order phase transition involved in the antiferromagnetic interaction is revealed at the level of static approximation \cite{Okuyama2018}.
In our approach, we propose belief propagation for part of the nonstoquastic Hamiltonian at the level of static approximation.
Building on the present analysis, we intend to employ the nontrivial form of the messages because of which we may have to deal with multipoint correlations.
Future studies beyond static approximation would reveal nontrivial aspects of the nonstoquastic Hamiltonian and bring out the potential of QA.

\begin{acknowledgments}
The authors would like to thank Masamichi J. Miyama, Shuntaro Okada, Shunta Arai and Shu Tanaka for fruitful discussions. 
The present work is financially supported by JSPS KAKENHI Grant No. 15H03699 and 16H04382, and Next Generation High-Performance Computing Infrastructures and Applications R\&D Program by MEXT, JST CREST (No. JPMJCR1402), ImPACT, and JST START.
\end{acknowledgments}
\bibliography{QAMF_ver1}
\end{document}